\documentclass[conference]{IEEEtran}

\usepackage[latin1]{inputenc}
\usepackage{epsfig}
\usepackage{graphicx}
\usepackage{amssymb}
\usepackage{amsmath}
\usepackage{subfigure}
\begin{document}

\title{Exact two-terminal reliability of some directed networks}

\author{\authorblockN{Christian~Tanguy\\}
\authorblockA{CORE/MCN/OTT\\France~Telecom~Division~R\&D\\%
38--40~rue~du~G\'{e}n\'{e}ral Leclerc,~92794~Issy-les-Moulineaux~Cedex~9,~France.\\%
christian.tanguy@orange-ftgroup.com}
}

\markboth{}{}

\maketitle

\begin{abstract}
The calculation of network reliability in a probabilistic context has long been an issue of practical and
academic importance. Conventional approaches (determination of bounds, sums of disjoint products algorithms,
Monte Carlo evaluations, studies of the reliability polynomials, etc.) only provide approximations when the
network's size increases, even when nodes do not fail and all edges have the same reliability $p$.

We consider here a {\em directed}, {\em generic} graph of {\em arbitrary} size mimicking real-life long-haul
communication networks, and give the exact, analytical solution for the two-terminal reliability. This
solution involves a product of transfer matrices, in which {\em individual} reliabilities of edges and nodes
are taken into account. The special case of identical edge and node reliabilities ($p$ and $\rho$,
respectively) is addressed. We consider a case study based on a commonly-used configuration, and assess the
influence of the edges being directed (or not) on various measures of network performance. While the
two-terminal reliability, the failure frequency and the failure rate of the connection are quite similar, the
locations of complex zeros of the two-terminal reliability polynomials exhibit strong differences, and
various structure transitions at specific values of $\rho$.

The present work could be extended to provide a catalog of exactly solvable networks in terms of reliability,
which could be useful as building blocks for new and improved bounds, as well as benchmarks, in the general
case.
\end{abstract}


\IEEEpeerreviewmaketitle

\section{Introduction}

\PARstart{N}{etwork} reliability has long been a practical issue, and will remain so for years, since
networks have entered an era of Quality of Service (QoS). IP networks, mobile phone networks, transportation
networks, electrical power networks, etc., have become ``commodities.'' Connection availability rates of
99.999\% are a goal for telecommunication network operators, and premium services may be deployed if the
connection reliability is close enough to one. Reliability is therefore a crucial parameter in the design and
analysis of networks.

The study of network reliability has led to a huge body of literature, starting with the work of Moore and
Shannon \cite{MooreShannon56}, and including excellent textbooks and surveys
\cite{Ball95,Barlow65,Colbourn87,Shier91,Shooman68,KuoZuo,Singh77}. In what follows, we consider a
probabilistic approach, in which the network is represented by an undirected graph $G = (V,E)$, where $V$ is
a set of nodes (also called vertices) and $E$ is a set of undirected edges (or links), each of which having a
probability $p_n$ or $p_e$ to operate correctly. Failures of the different constituents are assumed to occur
at random, and to be statistically independent events. Among the different measures of reliability, one may
single out the $k$-terminal reliability, namely the probability that a given subset $K$ of $k$ nodes ($K
\subset E$) are connected. The most common instances are the all-terminal reliability ${\rm Rel}_A$ ($K
\equiv E$) and the two-terminal reliability ${\rm Rel}_2(s \rightarrow t)$, which deals with a particular
connection between a source $s$ and a destination $t$. Both of them are affine functions of each $p_n$ and
$p_e$.

The sheer number of possible system states, namely $2^{|E|+|V|}$, clearly precludes the use of an
``enumeration of states'' strategy for realistic networks, and shows that the final expression may be
extremely cumbersome. Consequently, most studies have considered graphs with perfect nodes ($p_n~\equiv~1$)
and edges of identical reliability $p$; radio broadcast networks have also been described by networks with
perfectly reliable edges but imperfect nodes \cite{AboElFotoh89,Graver05}. It was shown early on --- see for
instance the discussion in \cite{Colbourn87,Welsh93} --- that the calculation of $k$-terminal reliability is
\#P-hard in the general case, even with the restricting assumptions that (i) the graph is planar (ii) all
nodes are perfectly reliable (iii) all edges have the same reliability $p$. All reliabilities are then
expressed as a polynomial in $p$, called the reliability polynomial.

The difficulty of the problem has stimulated many approaches: partitioning techniques \cite{Dotson79}, sum of
disjoint products \cite{Abraham79,Balan03,Heidtmann89,Rai95,RauzyChatelet03,Soh93}, graph simplifications
(series-parallel reductions \cite{MooreShannon56}, delta-wye transformations
\cite{Gadani81,Rosenthal77,Wang96}, factoring \cite{KevinWood85}), determination of various lower and upper
bounds to the reliability polynomial \cite{Ball95,Colbourn87,Beichelt89b,BrechtColbourn86,ScottProvan86},
Monte-Carlo simulations \cite{Fishman86,Karger01,Nel90}, and ordered binary decision diagram (OBDD)
algorithms \cite{Kuo99,Rauzy03,Yeh02,Yeh02conf}. Other decomposition methods have also been proposed
\cite{Carlier96,Bodlaender05,Galtier05}. The reliability polynomial has been extensively studied
\cite{Chari97,Colbourn93,Oxley02}, with the aim of finding general information from the structure of its
coefficients \cite{Chari97,Colbourn93} or the location of its zeros in the complex plane
\cite{BrownColbourn92}.

In recent years, the tremendous growth of Internet traffic has called for a better evaluation of the
reliability of connections in optical networks. Actual failure rates and maintenance data show that a proper
evaluation of two-terminal reliabilities must put node and edge equipments on an equal footing, i.e., both
edge (fiber links, optical amplifiers) and node (optical cross-connects, routers) failures must be taken into
account. The possibility of node failure has been considered in early papers \cite{AboElFotoh89,Hansler74}.
Adaptation of algorithms to include imperfect nodes has been addressed
\cite{Yeh02conf,Ke97,Netes96,Theologou91,Torrieri94}. In order to be realistic, different edge reliabilities
should be used too: for instance, the failure rate of optical fiber links is often assumed to increase with
their length.

In recent works, we have shown that the two- and all-terminal reliabilities can be exactly calculated for
recursive network architectures, where the underlying graphs are undirected and the edge/node reliabilities
arbitrary \cite{Tanguy06a,Tanguy06b,Tanguy06c}. The final expressions are products of transfer matrices, each
element of which is a multilinear polynomial of the individual edge or node reliabilities constituting the
`building block' (or `elementary cell') of the recursive graph.

%\vskip0.0cm
\begin{figure}[htb]
\centering
\includegraphics[scale=0.55]{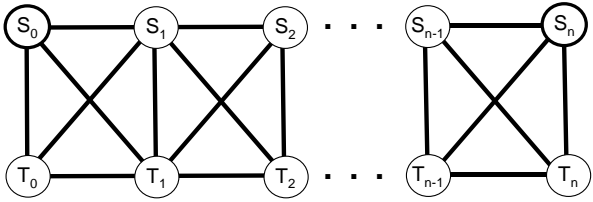}
\caption{General $K_4$ ladder. The source is always $S_0$, the destination is $S_n$ or $T_n$.} \label{Echelle
K4 generale}
\end{figure}

%\vskip0.0cm
\begin{figure}[htb]
\centering
\includegraphics[scale=0.5]{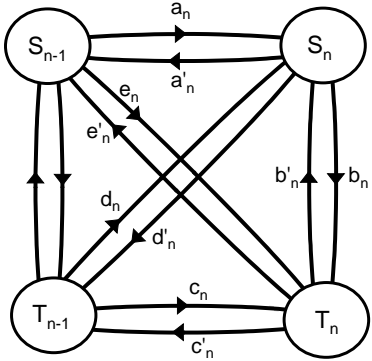}
\caption{Last building block of the directed $K_4$ ladder. Edges and nodes are identified by their
reliabilities.} \label{Brique de base de l'echelle K4}
\end{figure}

In this work, we show this general result holds for directed networks too. As an example, we calculate the
two-terminal reliability of the directed $K_4$ ladder displayed in Fig.~\ref{Echelle K4 generale}. This
network describes a common (nominal + backup paths) architecture, with additional connections between transit
nodes enabling the so-called ``local protection'' policy, which bypasses faulty intermediate nodes and/or
edges. By letting the individual node and edge reliabilities take {\em arbitrary} values, we actually do not
add to the complexity of the problem but make the internal structure of the problem more discernible. It is
then easier to fully exploit the recursive nature of the graph. The two-terminal reliability's exact
expression is a product of $5 \times 5$ transfer matrices, as in the undirected case \cite{Tanguy06c};
consequently, it can also be determined for an arbitrary size (length) of the network. If edges have the same
reliability $p$, and nodes the same reliability $\rho$, the two-terminal reliability can be expressed as a
sum over the eigenvalues of the unique transfer matrix, and its generating function is a rational fraction.
For large networks, the eigenvalue of highest modulus is, to all purposes, the scaling factor of the
asymptotic power-law behavior. The determination of the failure frequency and the failure rate of the
connection is then straightforward. Prompted by the nearly universal character of the Brown-Colbourn
conjecture \cite{BrownColbourn92}, we also address the location of complex zeros of the two-terminal
reliability polynomial and show that they (i) may be quite different for directed and undirected networks
(ii) exhibit structural transitions at various values of $\rho$.

Our aim is (i) to give a description of the methodology followed in the derivation of the final results, so
that researchers or engineers involved in reliability studies can use them, even in worksheet applications
(ii) emphasize again the importance of algebraic structures of the underlying graphs in the determination of
network reliability \cite{Shier91,Biggs93}.

Our paper is organized as follows. In Section~\ref{Graph decomposition}, we give the basic formula and
methodology used in the decomposition method \cite{Tanguy06c}, which must be adapted here for directed
graphs. In Section~\ref{Exact solution for the directed K4 ladder}, we give the exact solution for the
two-terminal reliability for the directed $K_4$ ladder (with source $S_0$ and destination $S_n$ or $T_n$). In
Section~\ref{Application}, we consider directed and undirected configurations of a special architecture. We
first give the generating functions of the two-terminal reliabilities when all edges and nodes reliabilities
are $p$ and $\rho$, respectively; we then derive very simple analytical expressions for ${\rm Rel}_2$. The
average failure frequency and the failure rate of the connection under consideration are then deduced. We
then show that the location of the complex zeros of ${\rm Rel}_2$ differ in the two configurations. We
conclude by proposing several directions in which the present results may be extended.

%\section{Graph decomposition}
\section{Graph decomposition}
\label{Graph decomposition}

The purpose of our method is to simplify the graph by removing links of the $n^{\rm th}$ ({\em last})
elementary cell of the network, namely the edges and nodes indexed by $n$, a procedure called pivotal
decomposition or deletion-contraction \cite{Colbourn87}. In the case of undirected graphs \cite{Tanguy06c},
we saw that if the end terminal $t$ (which can be regarded as perfect) is connected to node $u$ through edge
$e$, with respective reliabilities $p_u$ and $p_e$, then
\begin{eqnarray}
{\rm Rel}_2(G) & = & (1 - p_e) \, {\rm
Rel}_2(G \setminus e) +
p_e \, p_u \, {\rm Rel}_2(G \cdot e) \nonumber \\
& & + p_e \, (1-p_u) \, {\rm Rel}_2(G \setminus u) ,
\label{pivotaldecomposition}
\end{eqnarray}
where $G \setminus e$ and $G \setminus u$ are the graphs where $e$ or $u$ have been deleted, and $G \cdot e$
the graph where $t$ and $u$ have been merged through the ``contraction'' of $e$;
eq.~(\ref{pivotaldecomposition}) merely sums probabilities of disjoint events. Here, we must adapt this
decomposition to a directed graph. This is done very easily by first discarding all edges whose origin is $t$
and then applying eq.~(\ref{pivotaldecomposition}) to the remaining graph. This procedure, along with
standard series-parallel reductions, has to be repeated for the three secondary graphs in order to take
advantage of the structural recursivity of the graph. After a finite number of such reductions, we get
replicas of the original graph, albeit with one less elementary cell and with the $(n-1)^{\rm th}$ cell's
edge and node reliabilities possibly renormalized by those of the $n^{\rm th}$ cell, or set to either zero or
one. In order to ensure the existence of a recursion relation, the graph structure must be {\em closed} under
successive applications of eq.~(\ref{pivotaldecomposition}); it may initially require the use of extra edges
with symbolic reliabilities, so that all nodes of an elementary cell are connected pair-wise, even if such
links do not exist in the graph under consideration (the ``scaffolding principle''). At this point, a
recursion hypothesis is needed, giving for instance ${\rm Rel}_2(S_0 \! \rightarrow \! S_n)$ as a sum over
specific polynomials in the reliabilities indexed by $n$; these are often obvious from the $n=2$ value. Going
from $n-1$ to $n$ provides the transfer matrix linking the prefactors of the polynomials, because ${\rm
Rel}_2$ is an affine function of each component reliability; the (often trivial) $n=1$ case serves as the
initial condition of the recurrence.

%\section{Exact solution for the directed $K_4$ ladder}
\section{Exact solution for the directed $K_4$ ladder}
\label{Exact solution for the directed K4 ladder}

\begin{figure*}
\centerline{{\includegraphics[scale=0.55,angle=270]{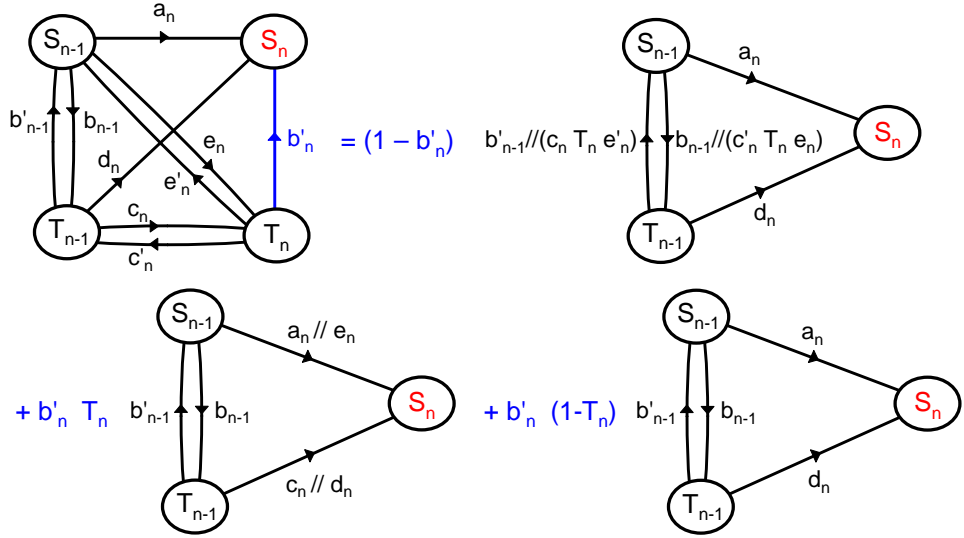}}} \caption{First step of the
decomposition. $a /\!/ b$ corresponds to $a$ and $b$ in parallel, and is therefore equal to $a + b - a \,
b$.} \label{Decomposition echelle K4}
\end{figure*}

Let us first illustrate this method by calculating ${\mathcal R}_n = {\rm Rel}_2(S_0 \! \rightarrow \! S_n)$
for the directed $K_4$ ladder shown in Fig.~\ref{Echelle K4 generale}; a more detailed view of the last
building block is displayed in Fig.~\ref{Brique de base de l'echelle K4}. Following the guidelines of the
preceding section, we remove the irrelevant edges, namely those indexed by $a'_n$, $b'_n$, and $d'_n$. The
first application of eq.~(\ref{pivotaldecomposition}) is represented in Fig.~\ref{Decomposition echelle K4},
where each prefactor is put in front of the associated secondary graph. Note that the three secondary graphs
are structurally identical, they differ by the actual values of the edge reliabilities only. Clearly, a
second application of eq.~(\ref{pivotaldecomposition}) provides two-terminal reliabilities, with $S_{n-1}$ or
$T_{n-1}$ as new endpoints. This should call for a similar decomposition of the two-terminal reliability
${\rm Rel}_2(S_0 \! \rightarrow \! T_n)$, after which we would get coupled recursion relations for the two
destinations $S_n$ and $T_n$. However, this is unnecessary because the two destinations are identical under
the permutations $a_n \! \leftrightarrow \! e_n$, $c_n \! \leftrightarrow \! d_n$ (and the corresponding
$a'_n \! \leftrightarrow \! e'_n$, etc.), and $S_n \! \leftrightarrow \! T_n$. It turns out that ${\mathcal
R}_n$ may be expressed as the sum of five polynomials in $a_n$, ..., $T_n$ (see below). This is also true for
${\rm Rel}_2(S_0 \! \rightarrow \! T_n)$, which leads us to a $10 \times 10$ transfer matrix (the
calculations are routinely performed by mathematical software). Because several lines of this matrix are
identical, regrouping terms actually allows to limit the transfer matrix's dimension to 5, as in the
undirected case \cite{Tanguy06c}. The value of ${\mathcal R}_{1}$, which can be easily calculated, leads to
\begin{equation}
{\mathcal R}_n = (1 \; 0 \; 0 \; 0 \; 0) \, M_n \, M_{n-1} \, \cdots M_1 \, M_0 \, \left(
\begin{array}{c}
1 \\
0 \\
0 \\
0 \\
0 \end{array} \right) ;\label{Rel2S0Snformelle}
\end{equation}
for ${\rm Rel}_2(S_0 \! \rightarrow \! T_n)$, the left vector should be $(0 \; 1 \; 0 \; 0 \; 0)$. The
transfer matrix $M_i$ is given by
%\small
\begin{equation}
M_i = \left(
\begin{array}{ccccc}
x_1 & x_2 & x_3 & x_4 & x_5 \\
x_6 & x_7 & x_8 & x_9 & x_{10} \\
x_{18} & x_{17} & x_{14} & -x_4-x_9 & -x_5-x_{10} \\
x_{20} & x_{19} & x_{15} & -x_9 & -x_{10} \\
x_{11} & x_{12} & x_{13} & -x_4 & -x_5
\end{array}
\right) ,
\end{equation}
%\normalsize
with
\begin{eqnarray*}
x_1 & = & {S_i}\,\left( {a_i} + {{b'}_i}\,{e_i}\,{T_i} - {a_i}\,{{b'}_i}\,{e_i}\,{T_i} \right)  , \\
x_2 & = & {S_i}\,\left( {d_i} + {{b'}_i}\,{c_i}\,{T_i} - {{b'}_i}\,{c_i}\,{d_i}\,{T_i} \right)  , \\
x_3 & = & {S_i}\,\left( {a_i}\,{d_i} + {a_i}\,{{b'}_i}\,{c_i}\,{T_i} -
     {a_i}\,{{b'}_i}\,{c_i}\,{d_i}\,{T_i} + {{b'}_i}\,{c_i}\,{e_i}\,{T_i} \right. \nonumber \\
     & & \left. -
     {a_i}\,{{b'}_i}\,{c_i}\,{e_i}\,{T_i} + (1 - a_i) \, {{b'}_i}\,(1 - c_i) \, {d_i}\,{e_i}\,{T_i} \right)  , \\
x_4 & = & \left( 1 - {a_i} \right) \,\left( 1 - {{b'}_i} \right) \,{{c'}_i}\,{d_i}\,{e_i}\,{S_i}\,
   {T_i} , \\
x_5 & = & {a_i}\,\left( 1 - {{b'}_i} \right) \,{c_i}\,\left( 1 - {d_i} \right) \,{{e'}_i}\,
   {S_i}\,{T_i} , \\
x_6 & = & \left( {e_i} + {a_i}\,{b_i}\,{S_i} - {a_i}\,{b_i}\,{e_i}\,{S_i} \right) \,{T_i} , \\
x_7 & = & \left( {c_i} + {b_i}\,{d_i}\,{S_i} - {b_i}\,{c_i}\,{d_i}\,{S_i} \right) \,{T_i} , \\
x_8 & = & \left( {c_i}\,{e_i} + {a_i}\,{b_i}\,{c_i}\,{S_i} + {a_i}\,{b_i}\,{d_i}\,{S_i} -
     {a_i}\,{b_i}\,{c_i}\,{d_i}\,{S_i} \right. \nonumber \\
     & & \left. - {a_i}\,{b_i}\,{c_i}\,{e_i}\,{S_i} + (1 - a_i) \, {{b}_i}\,(1 - c_i)\,{d_i}\,{e_i}\,{S_i}
     \right) \,{T_i} , \\
x_9 & = & {a_i}\,\left( 1 - {b_i} \right) \,{c_i}\,{{d'}_i}\,\left( 1 - {e_i} \right) \,{S_i}\,{T_i} , \\
x_{10} & = & {{a'}_i}\,\left( 1 - {b_i} \right) \,\left( 1 - {c_i} \right) \,{d_i}\,{e_i}\,{S_i}\,{T_i} , \\
x_{11} & = & \left( 1 - {a_i} \right) \,\left( 1 - {{b'}_i} \right) \,{e_i}\,{S_i}\,{T_i} , \\
x_{12} & = & \left( 1 - {{b'}_i} \right) \,{c_i}\,\left( 1 - {d_i} \right) \,{S_i}\,{T_i} , \\
x_{13} & = & \left( 1 - {{b'}_i} \right) \,\left( {a_i}\,{c_i} - {a_i}\,{c_i}\,{d_i} + {c_i}\,{e_i} -
     {a_i}\,{c_i}\,{e_i}\right. \nonumber \\
     & & \left.  + (1 - a_i) \,(1 - c_i) \,{d_i}\,{e_i} \right) \,{S_i}\,{T_i} , \\
x_{14} & = & \left( 1 - {{b'}_i} \right) \,\left( {a_i}\,{c_i} - {a_i}\,{c_i}\,{d_i} + {c_i}\,{e_i} -
     {a_i}\,{c_i}\,{e_i} \right. \nonumber \\
     & & \left. + (1 - a_i) \,(1 - c_i) \,{d_i}\,{e_i} \right) \,{S_i}\,{T_i} \\
& & - \left( {a_i}\,{b_i}\,{c_i} + {a_i}\,{b_i}\,{d_i} -
     {a_i}\,{b_i}\,{c_i}\,{d_i} + {c_i}\,{e_i} \right. \nonumber \\
     & & \left. - {a_i}\,{b_i}\,{c_i}\,{e_i} + (1 - a_i) \, {{b}_i}\,(1 - c_i)\,{d_i}\,{e_i} \right) \,{S_i}\, {T_i} , \\
x_{15} & = & \left( 1 - {b_i} \right) \,\left( {a_i}\,{c_i} + {a_i}\,{d_i} - {a_i}\,{c_i}\,{d_i} -
     {a_i}\,{c_i}\,{e_i} \right. \nonumber \\
     & & \left. + (1 - a_i) \,(1 - c_i) \,{d_i}\,{e_i} \right) \,{S_i}\,{T_i} , \\
x_{17} & = & - \left( c_i \, (b'_i + d_i - b'_i \, d_i) + b_i \, d_i \, (1 - c_i) \right) \, {S_i}\, {T_i} , \\
x_{18} & = & - \left( a_i \, (b_i + e_i - b_i \, e_i) + b'_i \, e_i \, (1 - a_i) \right) \, {S_i}\, {T_i} , \\
x_{19} & = & \left( 1 - {b_i} \right) \,\left( 1 - {c_i} \right) \,{d_i}\,{S_i}\,{T_i} , \\
x_{20} & = & {a_i}\,\left( 1 - {b_i} \right) \,\left( 1 - {e_i} \right) \,{S_i}\,{T_i} .
\end{eqnarray*}
For $i = 0$, we must set $a_0 = d_0 = 1$ and $c_0 = e_0 = 0$. These formulae apply to the most general
directed $K_4$ ladder, and we recover the undirected case by setting $a'_i = a_i$, $b'_i = b_i$, etc. A
missing edge or node is accounted for by setting the relevant reliability to zero, as will be seen in the
following section.

%\section{Application}
\section{Application}
\label{Application}

Let us apply the results of the preceding section to the architecture represented in Fig.~\ref{Echelles
Angele}. The calculations are straightforward, since we merely have to replace all nonexistent edges and
nodes by zero. Does the removal of these network elements drastically change the previous results? Actually,
no, even though in the directed case the dimension of the transfer matrix is reduced.

%\vskip0.0cm
\begin{figure}[htb]
\centering
\includegraphics[scale=0.45]{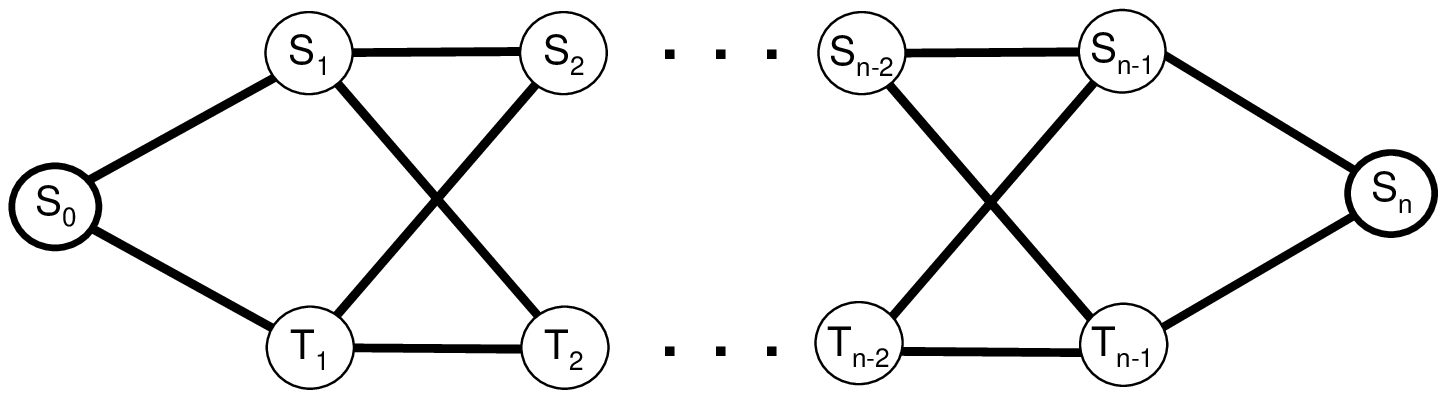}
\\
\vskip3mm
\includegraphics[scale=0.45]{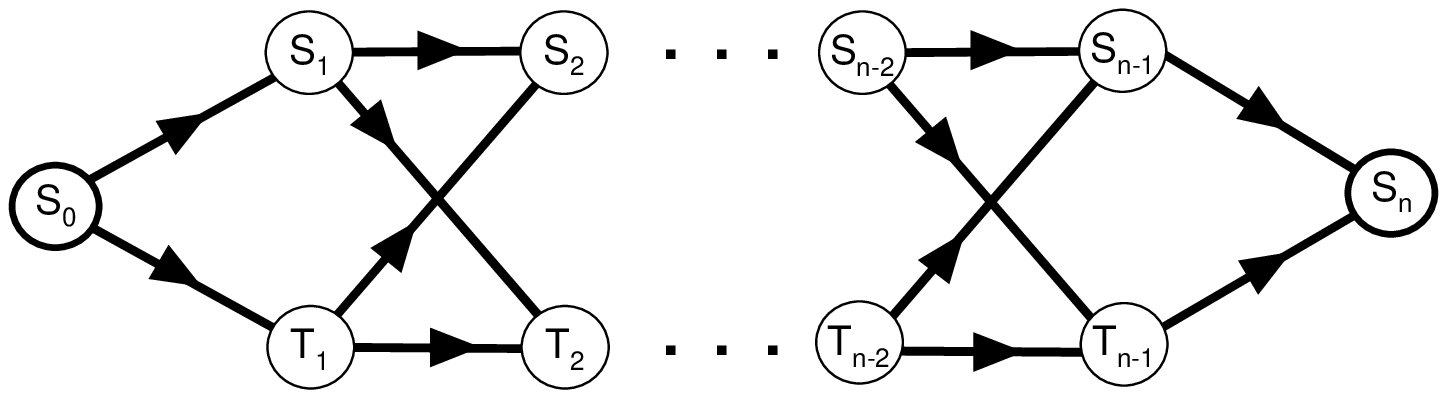}
\caption{Architecture discussed in this work (``\'{E}chelle Ang\`{e}le''). The source is $S_0$, the
destination is $S_n$ for both directed and undirected networks.} \label{Echelles Angele}
\end{figure}

\subsection{Transfer matrices}

In both directed and undirected configurations, we must set $T_0 = T_n = 0$ and $b_i =0$ ($0 \leq i \leq n$).
This does not change the dimension of the transfer matrix in the undirected case, which remains equal to 5.
However, for the directed configuration, the further simplification $a'_i = b'_i = c'_i = d'_i = e'_i = 0$ is
such that we can reduce the transfer matrix to a $3 \times 3$ one, namely
%\subsubsection{directed case}
\begin{equation}
\widetilde{M_i} = \left(
\begin{array}{ccc}
a_i \, S_i & d_i \, S_i & a_i \, d_i \, S_i \\
e_i \, T_i & c_i \, T_i & c_i \, e_i \, T_i \\
- a_i \, e_i \, S_i \, T_i & - c_i \, d_i \, S_i \, T_i & \chi_i \, S_i \, T_i
\end{array}
\right) ,
\label{M_iEchelleAngele}
\end{equation}
with $\chi_i = {a_i}\,{c_i} \, (1- {d_i}) \, (1- {e_i}) + {d_i}\,{e_i}\, (1  - {a_i} - {c_i})$, so that
\begin{equation}
\widetilde{{\mathcal R}}_n = (1 \; 0 \; 0) \, \widetilde{M}_n \, \widetilde{M}_{n-1} \, \cdots
\widetilde{M}_1 \, \widetilde{M}_0 \, \left(
\begin{array}{c}
1 \\
0 \\
0 \end{array} \right) .
\label{Rel2S0SnAngeleformelle}
\end{equation}

In many studies, edge reliabilities are considered identical to $p$ in order to provide clues to the general
behavior of the connection reliability, while nodes are viewed as perfect (i.e., their reliabilities are set
equal to 1). In this work, we keep imperfect nodes with identical reliability $\rho$. We thus have two
independent parameters to describe ${\rm Rel}_2(S_0 \! \rightarrow \! S_n)$, allowing us to better
distinguish the contributions of edges and nodes. Using only these two parameters implies that a unique
transfer matrix needs be considered. Equations~(\ref{Rel2S0Snformelle}) and (\ref{Rel2S0SnAngeleformelle})
show that the two-terminal reliability is given by the $n^{\rm th}$ power of this matrix. For instance,
eq.~(\ref{M_iEchelleAngele}) leads to
\begin{equation}
\widetilde{M}(p,\rho) = \left(
\begin{array}{ccc}
p \, \rho & p \, \rho & p^2 \, \rho \\
p \, \rho & p \, \rho & p^2 \, \rho \\
- p^2 \, \rho^2 & - p^2 \, \rho^2 & p^2 \, \rho^2 \, (2 - 4 \, p + p^2)
\end{array}
\right) . \label{M_iAngeleDirigeePRho}
\end{equation}

\subsection{Generating functions}
\label{generatingfunctions}

Because of the intrinsic recursion relation between successive powers of $\widetilde{M}(p,\rho)$, a similar
one should hold for the two-terminal reliability. The generating function formalism \cite{Stanley97} is a
useful way to store all necessary information in a very concise manner. It is defined by
\begin{equation}
{\mathcal G}(z) = \sum_{n=0}^{\infty} \, {\rm Rel}_2(S_0 \rightarrow S_n) \, z^n ,
\label{DefinitionFonctionGeneratrice}
\end{equation}
Its calculation is straightforward. Here, ${\mathcal G}(z)$ is necessarily a rational fraction of $z$, namely
\begin{equation}
{\mathcal G}(z) = \frac{{\mathcal N}(z)}{{\mathcal D}(z)} .
\end{equation}
Using eqs.~(\ref{Rel2S0SnAngeleformelle}) and (\ref{M_iAngeleDirigeePRho}) (for the directed configuration),
we compute the first values of ${\rm Rel}_2(S_0 \rightarrow S_n)$ for, say, $n$ equal to 2, 3, 4, etc., in
order to obtain the first terms of the expansion in $z$ of ${\mathcal G}(z)$. We then multiply this truncated
expression by the characteristic polynomial of the transfer matrix taken at $1/z$. The series expansion of
this product in the vicinity of $z=0$ leads to ${\mathcal N}(z)$. No determination of eigenvalues or
eigenvectors of the transfer matrix is needed.

Having described the method, we limit ourselves to the final, simplified expressions, which should be used
for $n \geq 2$, even though they are valid for $n = 1$ (both give the first-order term $p \, \rho^2 \, z$,
corresponding to the two-terminal reliability between the two nodes $S_0$ and $S_1$, connected by a single
edge).

\subsubsection{undirected case}

\begin{eqnarray}
{\mathcal N}_u(z) & = & \frac{\rho }{2\,\left( 1 - \rho  \right) } \;
\left(1 - p\,\left( 2 + 2\,p - 6\,p^2 + 3\,p^3 \right) \,{\rho }^2 \,z  \right. \nonumber \\
& & \left. \hskip-5mm + 2\,{\left( 1 - p \right) }^2\,\left( 2 - p \right) \,p^3\,\left( 1 - p + p^2 \right)
\,{\rho }^4  \,z^2\right) , \label{N_u}\\
{\mathcal D}_u(z) & = & 1 - p\,\rho \,\left( 2 + 2\,p\,\rho  - 6\,p^2\,\rho  + 3\,p^3\,\rho  \right)\,z \nonumber \\
& &  + 2\,\left( 1 - p \right) \,p^3\,{\rho }^3\, \left( 2 - 3\,p - 2\,p\,\rho  + 6\,p^2\,\rho \right. \nonumber \\
& & \left. \hskip2.8cm - 4\,p^3\,\rho  + p^4\,\rho  \right)\,z^2 \nonumber \\
& & - 4\,{\left( 1 - p \right) }^2\,\left( 2 - p \right) \,p^6\,\left( 1 - \rho  \right) \,{\rho }^5 \,z^3 \,
.
\label{D_u}
\end{eqnarray}

\subsubsection{directed case}

\begin{eqnarray}
{\mathcal N}_d(z) & = & \frac{\rho}{2} \; \left(1- p^2 \, \rho^2 \, (2 - 4 \, p + p^2) \, z \right) ,  \label{N_d} \\
{\mathcal D}_d(z) & = & 1  - p \, \rho \,\left( 2 + 2\,p\,\rho  - 4\,p^2\,\rho  + p^3\,\rho  \right) \,z
\nonumber \\
& & + 2\, p^3\,{\rho }^3 \, \left( 1 - p \right) \,\left( 2 - p \right) \, z^2  \, .
\label{D_d}
\end{eqnarray}

\subsection{Analytical expressions of the two-terminal reliabilities}
\label{Analytical reliabilities}

The two-terminal reliabilities are derived from the partial fraction decomposition of the associated
generating functions ${\mathcal G}_u(z)$ and ${\mathcal G}_d(z)$, because the eigenvalues of the transfer
matrix are the inverses of the generating function poles.

\subsubsection{undirected case}
Equations~(\ref{N_u}--\ref{D_u}) can be further simplified for perfect nodes ($\rho = 1$), because ${\mathcal
D}_u(z)$ is then of degree 2 in $z$, from which we find
\begin{equation}
{\mathcal R}^{(u)}_n (\rho = 1;n \geq 2) = a_+ \, \left(\zeta^{(u)}_+ \right)^n + a_- \, \left(\zeta^{(u)}_-
\right)^n ,
\end{equation}
with
\begin{eqnarray}
\zeta^{(u)}_{\pm} & = & \frac{p}{2}\,\left( 2 + 2\,p - 6\,p^2 + 3\,p^3 \pm {\sqrt{{\mathcal A}}} \right) ,
\label{Zeta+ undirected}\\
a_{\pm} & = &  \frac{ {\mathcal B} \pm
  \left( 1 - 4\,p + 2\,p^2 \right) \,{\sqrt{{\mathcal A}}}}{4\,{\left( 1 - p \right) }^2\,{\left( 1 - p
  + p^2 \right) }^2 \, {\sqrt{{\mathcal A}}}} \, , \label{A+ undirected} \\
{\mathcal A} & = & 4 - 8\,p + 36\,p^2 - 100\,p^3 + 128\,p^4 \nonumber \\
& & - 76\,p^5 + 17\,p^6 \, , \\
{\mathcal B} & = & 2 - 10\,p + 38\,p^2 - 59\,p^3 + 40\,p^4 - 10\,p^5.
\end{eqnarray}
For $\rho < 1$, we would get a sum over three eigenvalues, the analytical expression of which would only be
more cumbersome (it will not be given here).

\subsubsection{directed case}

Even if $\rho \neq 1$, ${\mathcal D}_d(z)$ is of degree 2 in $z$. We have again two eigenvalues, so that
\begin{equation}
{\mathcal R}^{(d)}_n (n \geq 2) = \alpha_+ \, \left(\zeta^{(d)}_+ \right)^n + \alpha_- \, \left(\zeta^{(d)}_-
\right)^n ,
\end{equation}
with
\begin{eqnarray}
\zeta^{(d)}_{\pm} & = & \frac{p \, \rho}{2} \; \left( 2 + p \, \rho \, (2 - 4 \, p + p^2)
\pm \sqrt{{\mathcal A}'} \right) , \label{Zeta+ directed}\\
\alpha_{\pm} & = & \frac{\rho}{4} \; \left( 1 \pm  \frac{2 - p \, \rho \, (2 - 4 \, p + p^2)}
{\sqrt{{\mathcal A}'}} \right) , \label{Alpha+ directed}\\
{\mathcal A}' & = & 4 - 4 \, p \, \rho \, (2 - 2 \, p + p^2) \nonumber \\
& & + p^2 \, \rho^2 \, (2 - 4 \, p + p^2)^2 \, .
\end{eqnarray}

\begin{figure}[thb]
\centering
\includegraphics[scale=0.8]{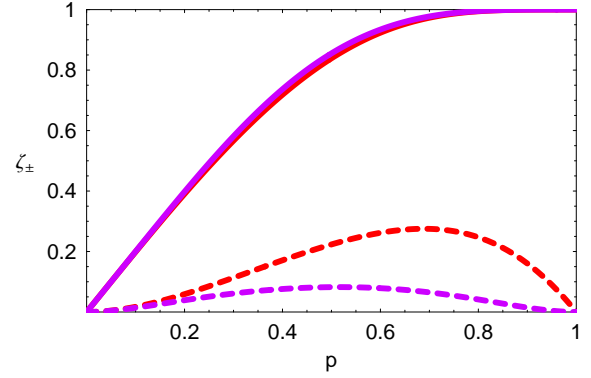}
\caption{Variation of $\zeta_{+}$ (full line) and $\zeta_{-}$ (dashed line)
with $p$ for the directed (red curve) and undirected (purple curve)
architectures of Fig.~\ref{Echelles Angele} with perfect nodes.}
\label{Variation valeurs propres}
\end{figure}

$\zeta^{(u)}_{\pm}$ and $\zeta^{(d)}_{\pm}$ are displayed in Fig.~\ref{Variation valeurs propres} for perfect
nodes. Obviously, $\zeta_+$ is nearly always much larger than $\zeta_-$, especially when $p$ lies in the
vicinity of 1. $\zeta^{(u)}_{+}$ and $\zeta^{(d)}_{+}$ are nearly equal over the whole range $0 \leq p \leq
1$, while the second eigenvalue is much larger in the directed case. Still, as $n$ grows, the contribution of
the second eigenvalue should vanish so that the two-terminal reliability exhibits an asymptotic power-law
behavior ${\mathcal R}_n \propto \zeta_+^n$, the scaling factor being $\zeta_+$, the eigenvalue of largest
modulus. Even for $n \approx 10$, this asymptotic limit would already be a good approximation.

%\vskip0.0cm

\subsection{Average failure frequency and failure rate}
\label{Average failure rates}

Steady-state system availability $A$ and failure frequency $\overline{\nu}$ are important performance
measures of a repairable system, from which other key parameters such as the mean time between failures,
average failure rate, Birnbaum importance, etc. may be deduced \cite{Shooman68,KuoZuo,Singh77}. The first
calculations of steady-state failure frequencies were based on the inclusion-exclusion principle, with
adequate failure and repair rates (more generally, the inverses of the mean down and up times) were
attributed to each term of the expansion \cite{Singh74,Singh75}. Several papers have since provided a few
simple recipes, describing how $\overline{\nu}$ and the system failure rate $\overline{\lambda} =
\overline{\nu}/A$ can then be derived \cite{Schneeweiss81,Shi81,Amari00,Chang04}. All these formal
calculations boil down to a simple fact: the failure frequency may be derived from $A$ by the application of
a linear differential operator \cite{Schneeweiss83,Hayashi91,ADVCT06}:
\begin{equation} \overline{\nu} =  \sum_i \lambda_i \,
p_i \, \frac{\partial A}{\partial p_i} = \sum_i \mu_i \, q_i \, \, \frac{\partial U}{\partial q_i} \, ,
\label{nuGenerale}
\end{equation}
where $U = 1 - A$ is the total unavailability, $p_i$ the availability, $q_i = 1 - p_i$ the unavailability,
$\lambda_i$ the failure rate, and $\mu_i$ the repair rate of equipment $i$. The expressions obtained in the
preceding sections ($A \equiv {\mathcal R}_n$) make such calculations straightforward by the application of
the linear differential operator to each transfer matrix. Good estimates of what happens for large networks
($n \gg 1$) may be obtained by considering that all links have availability $p$ and failure rate $\lambda$
(assuming nodes are perfect, to keep the discussion simple). If ${\mathcal R}_n \approx a_+ \, \zeta_+^n$,
the average failure rate $\overline{\lambda}_n$ is then given by
\begin{equation}
\overline{\lambda}_n = \frac{\overline{\nu}_n}{{\mathcal R}_n} \approx \lambda \; \left( \frac{\partial \ln
a_+}{\partial \ln p} + n \, \frac{\partial \ln \zeta_+}{\partial \ln p} \right) \, .
\end{equation}
%The analytical expressions of $\overline{\lambda}_n$ are easily obtained from eqs.~(\ref{Zeta+ undirected}--\ref{A+ undirected}) and (\ref{Zeta+ directed}--\ref{Alpha+ directed}).

%\vskip0.0cm
\begin{figure}[thb]
\centering
\includegraphics[scale=0.8]{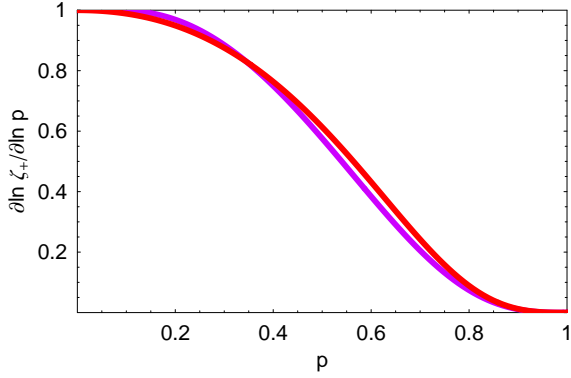}
\caption{Variation of $\partial \ln \zeta_+/\partial \ln p$ with $p$ for the directed (red curve) and
undirected (purple curve) architecture of Fig.~\ref{Echelles Angele} with perfect nodes.} \label{Taux Panne
Echelle Angele}
\end{figure}

The ``effective'' network can therefore be seen as a series network of $n$ components, having each an
availability equal to $\zeta_+$ and a failure rate $\frac{\partial \ln \zeta_+}{\partial \ln p} \, \lambda$.
From Fig.~\ref{Variation valeurs propres} we could expect that the asymptotic failure rate for our case study
would be roughly similar for directed and undirected configurations, over the whole range $0 \leq p \leq 1$.
This is indeed observed in Fig.~\ref{Taux Panne Echelle Angele}.

\subsection{Zeros of the reliability polynomials}

The structure of the different reliability polynomials may be understood by studying the locations of their
zeros in the complex plane. Such a study has been fruitfully performed for chromatic polynomials
\cite{Biggs72,Biggs01,Salas01}. In reliability studies, a search for general properties of the all-terminal
reliability polynomial ${\rm Rel}_A(p)$ \cite{Chari97,Colbourn93,Oxley02} has brought the Brown-Colbourn
conjecture \cite{BrownColbourn92}, according to which all the zeros could be found in the region $|1-p| < 1$.
Although valid for series-parallel graphs, this remarkable conjecture does not strictly hold in the general
case (but not by far) \cite{Royle04}. ${\rm Rel}_A(p)$ is linked to the Tutte polynomial, a graph invariant.
It has also been studied extensively by Chang and Shrock for various recursive families of graphs
\cite{Chang03}, who give the limiting curves where all the zeros converge.

As $n$ grows, the number of complex zeros of the reliability polynomial increases. Because of the matrix
transfer property, there is a recursion relation between reliability polynomials corresponding to successive
values of $n$. The general problem has been treated by Beraha, Kahane, and Weiss \cite{Beraha78}. It may be
understood in the following, simplifying way: if the reliability polynomial is of the form $\sum_i \alpha_i
\, \lambda_i(p)^n$ (where $\lambda_i$ are the eigenvalues of the recurrence), then at large $n$, only the two
eigenvalues of greater modules, say $\lambda_1$ and $\lambda_2$, will prevail, so that the reliability
polynomial will vanish when $|\lambda_1(p)| = |\lambda_2(p)|$. This equality defines a set of curves in the
complex plane, where all zeros should accumulate in the $n \rightarrow \infty$ limit. A detailed discussion
of the convergence to the limiting curves may be found in \cite{Salas01}.

The location of the zeros of ${\rm Rel}_2(p)$ in the complex plane is also worth investigating, even though
${\rm Rel}_2(p)$ is not a graph invariant. The new twist lies in the extra parameter at our disposal, the
node reliability $\rho$, which has a deep impact on the curves to which the zeros of ${\rm Rel}_2(p)$
converge as $n \rightarrow \infty$. Actually, structural changes occur at critical values of $\rho$, which
can be deduced from the expressions of the eigenvalues.

We have displayed in Figs.~\ref{zeros non oriente} and \ref{zeros oriente} the location of complex zeros of
the two-terminal reliability polynomials from our case study (the numerical values have been obtained by
using {\sc Mathematica}). The structures of these zeros are quite different for the directed and undirected
cases. In both configurations, parts of the real positive axis are indeed the limiting ``curve''. It turns
out that in the undirected case, there is no such real line segment for exactly $\rho_{c_1} = \frac{8}{9}$
(the sole intersection of the right-most curve with the real axis lies at $p = \frac{3}{2}$). We shall not
detail the other critical values for $\rho$. In the directed case, the segment gets closer and ``punctures''
the curve on the right half-plane for $\rho_{c_2} \approx 0.51242$ (a root of a polynomial of degree 10). As
$\rho$ further decreases, both structures look like circles plus an extra line segment, which expands as
$\rho^{-1/3}$. We limit ourselves to the asymptotic values of $|p_{\rm circle}|$ (the circle) and $p_{\pm}$
(the endpoints of the segment on the real positive axis). Note the expansion rates are distinct, too.

\subsubsection{undirected case}
\begin{eqnarray}
p_{\pm}^{(u)} & \rightarrow & \left( \frac{\sqrt{17}-3}{2 \, \rho}\right)^{\frac{1}{3}} \pm \sqrt{\frac{34 -
2 \,
\sqrt{17}}{153}} \, \left( \frac{\sqrt{17}-3}{2 \, \rho}\right)^{\frac{1}{6}} \nonumber \\
& & + \frac{2}{51} \, (23  - \sqrt{17}) \, , \\
|p_{\rm circle}^{(u)}| & \rightarrow & \left( \frac{\sqrt{17}-3}{2 \,
\rho}\right)^{\frac{1}{3}} \, .
\end{eqnarray}

\subsubsection{directed case}
\begin{eqnarray}
p_{\pm}^{(d)} & \rightarrow & \left( \frac{2}{\rho}\right)^{\frac{1}{3}} \pm \frac{2}{3} \, \left(
\frac{2}{\rho}\right)^{\frac{1}{6}} + \frac{4}{3}  \, , \\
|p_{\rm circle}^{(d)}| & \rightarrow & \left(
\frac{2}{\rho}\right)^{\frac{1}{3}} \, .
\end{eqnarray}

{
\begin{figure}[thb]
\centering
\includegraphics[scale=0.80]{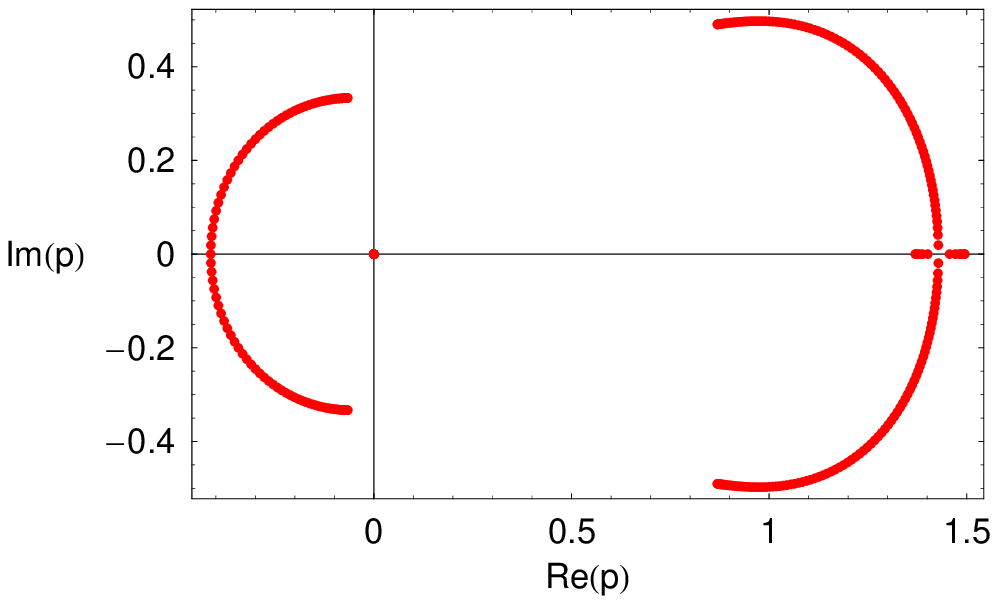}
\\
\includegraphics[scale=0.80]{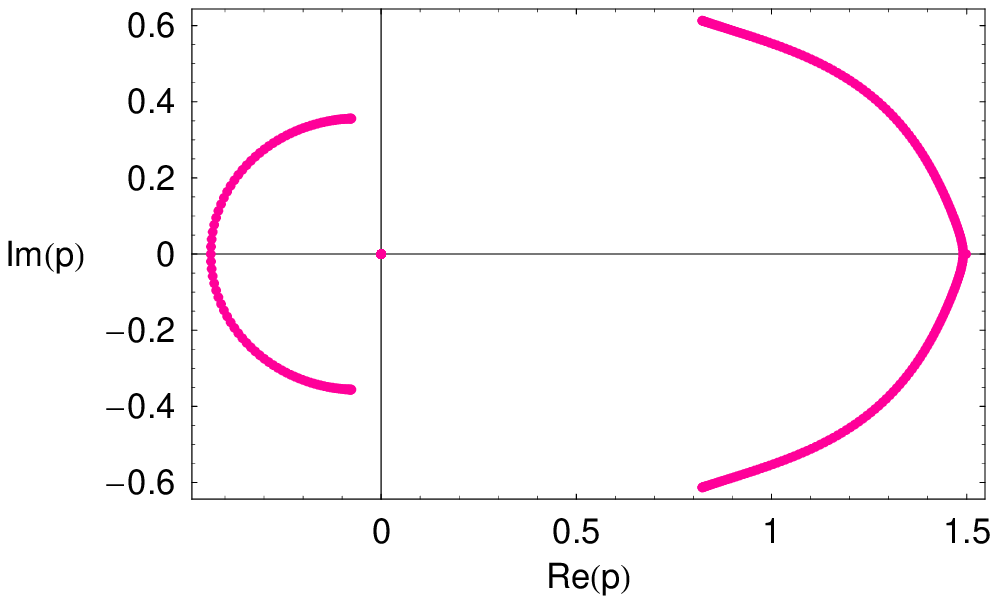}
\\
%\includegraphics[scale=0.80]{ZerosEchelleAngelen100rho0.5.eps}
%\\
\includegraphics[scale=0.80]{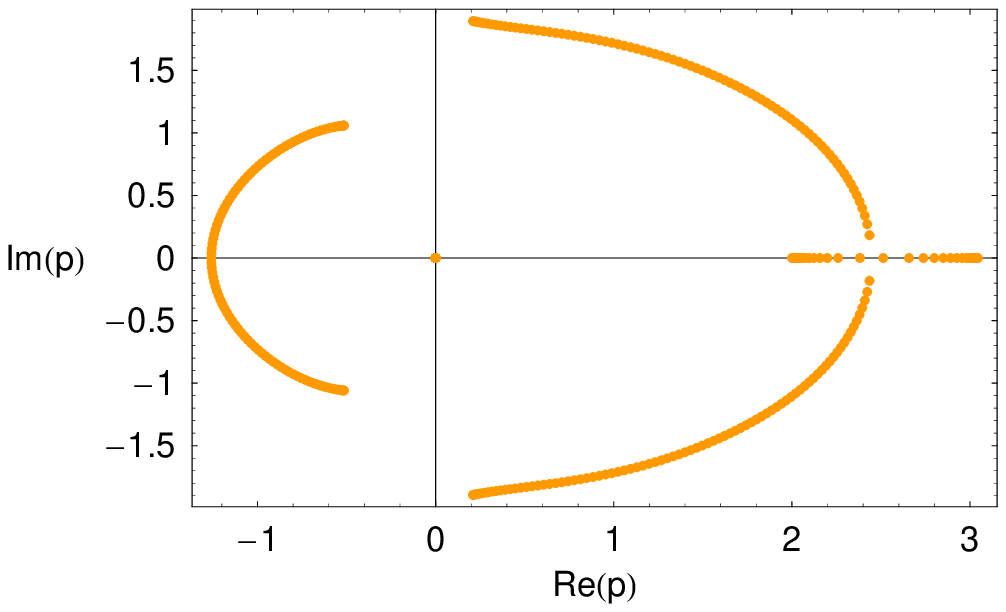}
\caption{Location of complex zeros for the undirected ladder, with $n=100$ and $\rho$ equal to 1, 0.9, and
0.1 (from top to bottom).} \label{zeros non oriente}
\end{figure}

\begin{figure}[thb]
\centering
\includegraphics[scale=0.80]{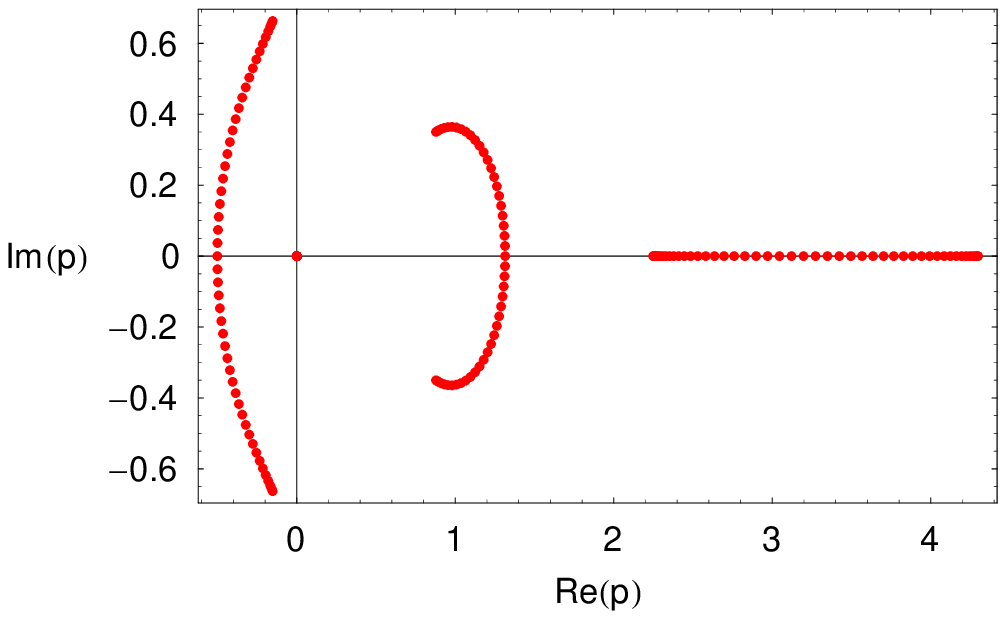}
\\
\includegraphics[scale=0.80]{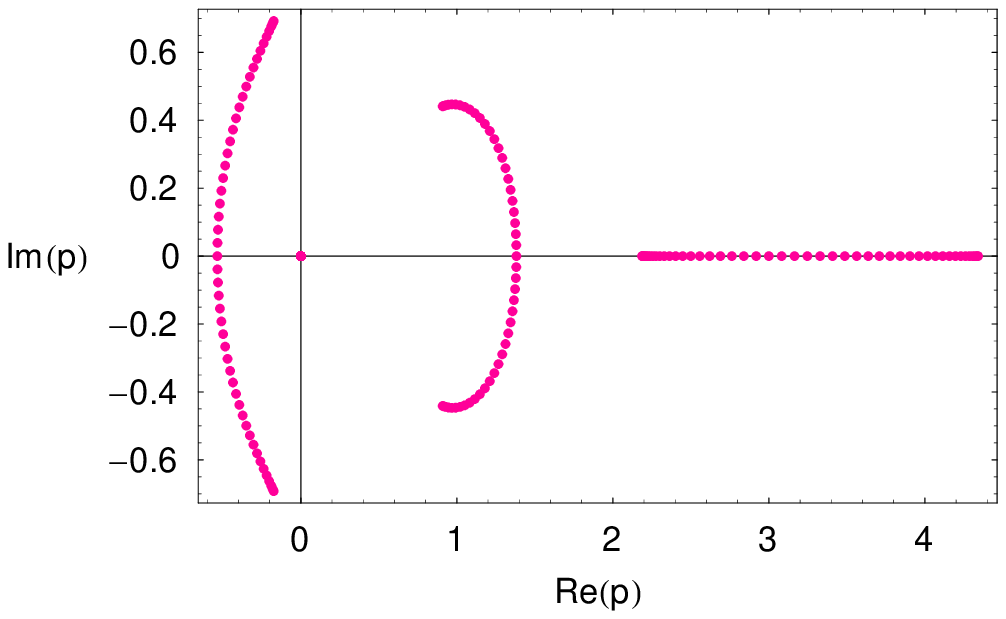}
\\
%\includegraphics[scale=0.80]{ZerosEchelleAngeleOrienteen50rho0.5.eps}
%\\
\includegraphics[scale=0.80]{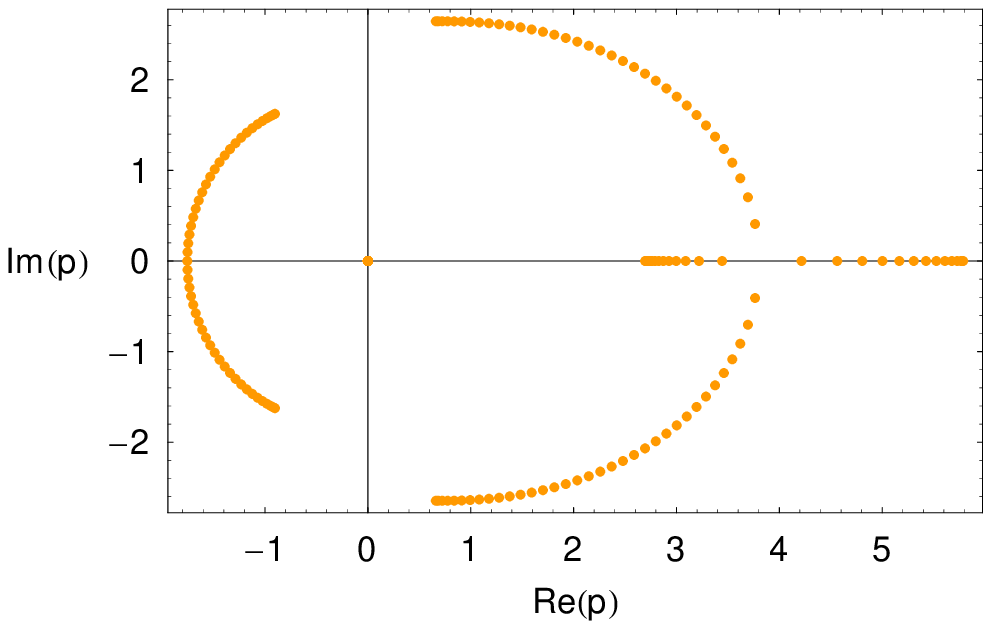}
\caption{Location of complex zeros for the directed ladder, with $n=50$ and $\rho$ equal to 1, 0.9, and 0.1
(from top to bottom).} \label{zeros oriente}
\end{figure}
}

\section{Conclusion and outlook}
\label{Conclusions}

The two-terminal reliability of directed networks may also be expressed by a
product of transfer matrices, in which each edge and node reliability is
exactly taken into account. The size of the transfer matrix should of course
increase with the network's ``width''. This general result could be extended to
the all-terminal reliability with nonuniform links \cite{Sokal05}. We can now
go beyond series-parallel simplifications and look for new (wider) families of
exactly solvable, meshed architectures that may be useful for general
reliability studies (as building blocks for more complex networks), for the
enumeration of self-avoiding walks on lattices, and for directed percolation
with imperfect bonds {\em and} sites. Since the true generating function is
itself a rational fraction, Pad\'{e} approximants could provide efficient upper
or lower bounds for these studies. Moreover, individual reliabilities can be
viewed as average values of random variables. Having access to {\em each} edge
or node allows the introduction of disorder or correlations in calculations
\cite{Coit04}.

\section*{Acknowledgment}
The author would like to thank Ang\`{e}le Phu for the architecture described in
this work, Éric Gourdin, James Roberts, and Sara Oueslati for helpful
discussions, and Olivier Klopfenstein and Beno\^{i}t Lardeux for very useful
suggestions.

\end{document}